\documentclass[doublecol]{epl2} 

\usepackage{amsmath}
\usepackage{amsthm}
\usepackage{graphicx}
\usepackage{dcolumn}
\usepackage{bm}

\newcommand{\ep}{\epsilon}

\newcommand{\obs}{{\cal O}}

\newtheorem{theorem}{Theorem} 
\theoremstyle{definition}

\theoremstyle{remark}

\title{On the robustness of $q$-expectation values and R\'enyi entropy}
\shorttitle{Robustness of $q$-expectations} 

\author{Rudolf Hanel\inst{1} \and Stefan Thurner\inst{1,3} \and Constantino Tsallis\inst{2,3}}
\shortauthor{R. Hanel  \etal}

\institute{                    
  \inst{1} Complex Systems Research Group, HNO, Medical University of Vienna, W\"ahringer G\"urtel 18-20, A-1090, Austria\\
  \inst{2} Centro Brasileiro de Pesquisas F\'isicas, Rua Xavier Sigaud 150, 22290-180 Rio de Janeiro-RJ, Brazil\\
  \inst{3} Santa Fe Institute, 1399 Hyde Park Road, Santa Fe, NM 87501, USA
}

\abstract{
We study  the robustness of  functionals of  probability distributions such as the R\'enyi and nonadditive $S_q$ entropies, 
as well as the  $q$-expectation values under small variations of the distributions. 
We  focus on three  important types of distribution functions, namely (i) continuous bounded (ii) discrete 
with finite number of states, and (iii) discrete with infinite number of states. The physical concept of robustness is 
contrasted with the mathematically stronger condition of  stability and Lesche-stability for functionals. 
We explicitly demonstrate that, in the case of continuous distributions, once   
unbounded distributions and those leading to negative entropy  are excluded, both R\'enyi and nonadditive $S_q$ 
entropies as well as the $q$-expectation values are robust. 
For the discrete  finite case, the  R\'enyi and nonadditive $S_q$ entropies and the $q$-expectation values are robust. 
For the infinite discrete case, where both R\'enyi entropy and $q$-expectations are known 
to violate Lesche-stability and stability respectively, we show that one can nevertheless 
state conditions which guarantee physical robustness.  
}

\begin{document}

\maketitle
 
\section{Introduction}

Thermostatistical  quantities such as entropy are expressed as functionals 
of probability distributions. For these quantities to be physically meaningful they should not 
change drastically if the underlying distribution functions are slightly changed. In practical terms, 
the unavoidable experimental uncertainty in determining the distribution function should not cause the 
thermodynamical quantities to fluctuate wildly, or even diverge. It is therefore of elementary interest  
to clarify and check that thermodynamical quantities are robust under small variations of the distribution functions. 
We define probability distributions on a set of $W$ discrete states, $p=\{p_i\}_{i=1}^W$. 
Let us denote a variation 
by $p_i'=p_i+\delta p_i$, the $L_1$ distance being $|| p-p'||_1 = \sum_{i=1}^W|p_i-p_i'|$. 

In this context, almost three decades ago, Lesche has introduced a definition of stability of 
functionals \cite{lesche}. There a functional $Q[p]$  (e.g., an entropic form) is called {\em stable}  
({\em Lesche-stable}, as often referred to nowadays) if, for every $\epsilon$, one can 
find  a $\delta$ 
such that for {\em all}  $W$ and for {\em all} $p$ and $p'$ one has
\begin{equation}
  ||p-p'||_1<\delta   \,\,  \Rightarrow  \,\,   \frac{ \left|Q[p]-Q[p']\right| }{Q^{\rm max }}<\ep    \quad.
\label{lesche} 
\end{equation}
Here $Q^{\rm max}$ is the maximum of the functional. 
Lesche could show that, under this strict definition of stability, R\'enyi entropy is unstable. Indeed,  
he could find  examples for $p$ and $p'$  for which --- by taking the $W\to \infty$ limit --- 
Lesche-stability is violated \cite{lesche}.  
There it is also mentioned that the $L_1$-norm has to be used because in this norm 
some relevant statistical quantities become independent of the number of states $W$.
Taking 
the $W\to\infty$ limit is essential. 
If one can show that, in the $W\to \infty$ limit,  stability is violated, this 
implies that for some finite $W$ violation  is already emerging,
i.e. the bound $\ep$ in Eq. (\ref{lesche}) gets violated for specific distributions $p$ and $p'$
with $||p-p'||_1<\delta$.
Hence,
the condition is 
not true for {\em all} $W$, the functional  thus 
is Lesche-unstable. 
Lesche-stability has been used lately to  analyze  the stability of various 
generalized entropies. 
The use of  entropies on that basis was suggested as a new validity criterion 
\cite{lesche,abe00,russe,abe04}. 
It is therefore not surprising that it   
has  occasionally lead to some confusion, see e.g.  \cite{lesche04,yamano}.
The discussion of Lesche-stability has also been extended to other entropies \cite{kaniadakis,curado}. 
If one does not divide by $Q^{\rm max }$, Eq. (\ref{lesche}) becomes the  traditional continuity condition  
for a functional and the stability criterion becomes a notion of uniform continuity. 

The requirement that a functional $Q[p]$ should only be considered stable if 
Eq. (\ref{lesche})  holds for {\em all } $p$ and $p'$ and $W$ uniformly,  
is  unnecessarily strict  for physical systems. In
a physical context it is reasonable to call a functional $Q[p]$ {\em physically robust} 
if it is continuous on the domain of {\em physically  admissible} distributions $p$. 
In the case of continuous distribution functions, admissible  requires that the  corresponding 
Boltzmann Gibbs (BG) entropy 
($-\int p \ln p$) is 
positive\footnote{In the continuum, entropy functionals such as the BG, R\'enyi and others are well known to 
become negative for distributions which include too narrow peaks, a situation which typically 
corresponds to the  low temperature limit, where the quantum nature of physical systems 
{\em must} be taken into account.}.  
In other words, for 
physical situations it is sufficient to ensure robustness of 
functionals, {\em not} necessarily their stability.\footnote{If a functional is stable 
it is always safe to use. Inversely, instability points at the fact that the 
domain of safe usage is limited. 
Robustness is never used in the sense of trajectories or attractors.} 
Although unbounded physical distributions exist, we are not focusing 
on them here. 
For all other physically admissible distribution functions robustness is guaranteed.
For finite $W$ any probability distribution is admissible. 

The discussion of stability and robustness is not restricted to entropies
\cite{abe00,russe,abe04,lesche04,yamano,kaniadakis,curado}, but also to other 
quantities such as the $q$-expectation values, which naturally 
occur in the context of  formalisms using generalized entropic functionals  
\cite{tsallis88,tsallismendesplastino,naudts_ent,abeescort}. 
The $q$-expectation values (i.e., standard expectation values with the so-called escort distribution \cite{beck}, 
proportional to $p^q$) naturally appear  
in differential thermodynamic relations whenever the probability distribution presents power-law behavior.
This can be illustrated with the $q$-exponential function, $e_q(x)\equiv [1+(1-q)x]^{\frac{1}{1-q} }$, 
which, for $q>1$,  asymptotically decays like a power-law. Indeed, whenever 
one takes  derivatives 
of usual expectation values escort expectation values can not be avoided, 
since the  exponent $q$ emerges due to $d e_q(x) /dx = [e_q(x)]^q$. 
For instance, normalization of the typical $q$-exponential distribution $\rho(\ep)=e_q(-\alpha-\beta\ep)$,
where $\beta$ is the inverse temperature and $\alpha$ a normalization constant,
requires $1=\int_0^\infty d\ep\,e_q(-\alpha-\beta\ep)$. A simple calculation shows that the derivative 
$d\alpha/d\beta=-\int d\ep \rho^q(\ep)\ep/\int d\ep \rho^q(\ep)$, which is exactly the escort expectation of $\ep$.
Another aspect can be illustrated with unimodal distributions. 
For example, if one has a $q$-Gaussian distribution \cite{prato}, its width can be characterized 
by the $(variance)^{1/2}$ as long as $q < 5/3$. 
This is not true anymore if $q \geq 5/3$ (e.g. for $q=2$, which corresponds to 
the celebrated Cauchy-Lorentz distribution)  since the variance {\em diverges}. In all cases, however, 
we can characterize the width by the inverse of the maximal 
value of the distribution. It happens that this inverse 
scales like the  $(q-variance)^{1/2}$.

Recently it was shown,  using Lesche's two explicit examples for $p$ and $p'$   \cite{lesche}, 
that $q$-expectation values are unstable on discrete  infinite distributions \cite{abe}. 
The first example  corresponds to $0<q<1$, the  second one to $q>1$: 
\\
{\bf Example} (1): $0<q<1$
\begin{equation}
p_i=\delta_{i\,1}\quad,\quad p_i'=\left(1-\frac{\delta}{2}\frac{W}{W-1}\right)p_i+\frac{\delta}{2}\frac{1}{W-1}
\label{case1}
\end{equation}
{\bf Example} (2): $q>1$
\begin{equation}
p_i=\frac{1}{W-1}\left(1-\delta_{i\,1}\right) \,\, ,\,\, p_i'=\left(1-\frac{\delta}{2}\right)p_i+\frac{\delta}{2}\delta_{i\,1}\quad. 
\label{case2}
\end{equation}
Here $||p-p'||_1=\delta$, for any value of $W$. 
Specifically, in \cite{abe}, instability  was shown for  
the $q$-expectation of an observable $\obs=\{\obs_i\}_{i=1}^W$  on the discrete 
index set $\tilde I=\{1,\dots,W\}$, i.e., the expectation,  with $q\neq1$, of $ Q[p]=\sum_i P_i^{(q)}\obs_i$, 
where the escort distribution is given by 
\begin{equation}
P_i^{(q)}= \frac{p_i^q}{\sum_{j=1}^W p_j^q} \,.
\label{escort}
\end{equation}

For both examples  
$\lim_{W\to\infty}| Q[p]-Q[p']|=|\bar\obs-\obs_1|$, where $\bar\obs \equiv \lim_{W\to\infty}W^{-1}\sum_i \obs_i$, which
proves instability when $\obs$ and $K$ are chosen such that $| \bar \obs-\obs_1|>K>0$, \cite{abe}.
This implies that $q$-expectations are not {\em uniformly} continuous functionals in the $\lim W\to\infty$. 
It was concluded in \cite{abe}  that the instability of the $q$-expectation value is the general situation,  thus
suggesting to re-think the use of $q$-expectation values in nonextensive statistical mechanics.
While the result in \cite{abe} is correct  in the strict sense of stability used there, 
this does {\em not} imply that $q$-expectation values are not robust either  on finite sets -- as will be shown here --
or for  continuous variables with the mentioned  physical admissibility and boundedness conditions \cite{hanel08}. 
Therefore the final conclusion drawn in \cite{abe} that the $q$-expectation is {\em  in general} 
unstable under small variations 
of the probability distributions  does {\em not} hold for physically relevant cases such as   
continuum distributions and discrete distributions on finite support. 
Even for a discrete infinite support, robustness is verified, as it will be shown, for paradigmatic physical distributions.
Robustness in the above sense is sufficient for virtually all practical physical purposes. 
In other words the requirements of boundedness and positive entropy exclude 
the pathological cases of singular distributions  
and singular variations. The examples used in \cite{abe} 
are representatives of such  pathological cases.  

In this contribution we  primarily discuss stability and robustness of $q$-expectation values and 
R\'enyi entropy  for three types of support for distribution functions, the continuous, discrete finite, 
and discrete infinite. 
For the continuous case, under the requirement of physically admissible and bounded 
probability distributions and variations, we show the robustness of $q$-expectation values and 
R\'enyi entropy. We then show that the theorems 
used in \cite{hanel08}  allow to prove robustness for {\em all} finite discrete sets. 
Even though it is not possible to immediately use the theorems to make statements 
about  infinite discrete sets, which have been shown to be unstable for R\'enyi entropy and 
the $q$-expectation value (for $q\neq1$) \cite{lesche,abe},  we show how the theorems can be used 
to derive restrictions so that 
robustness can be accomplished there as well. 
We finally discuss the situation for $S_q$ entropy for continuous and discrete finite 
probability distributions. A discussion on the distinction of  discrete finite  and infinite 
cases has been presented on numerical grounds in \cite{tsallis04}.  It is known that the nonadditive entropy $S_q$ on 
discrete infinite distributions is robust because it is  Lesche-stable \cite{abe00}.

\section{Stability criteria for the $q$-expectation value for admissible continuous distribution functions}

To make the paper self-contained we first review the stability criteria for the continuum case
as discussed in two theorems in \cite{hanel08}. 
These two theorems determine  the robustness  criteria
for $q$-expectation values in the continuum. 
These theorems will be used below to show that not only the two examples of 
Lesche used in \cite{abe} are robust on finite sets, 
but that this is the case for {\em all} distribution functions on finite sets.  

For notation, in the continuum, the escort distribution reads  
$P^{(q)}(x)\equiv \frac{\rho(x)^q }{ \int dx' \rho(x')^q } $, where $\rho$ denotes a continuous 
probability distribution. 
The expectation value of an observable $\obs(x)$ under this measure  is   
$\tilde Q[\rho] = \int dx  P^{(q)}(x)\obs(x)$, and its 
total variation reads $\delta\tilde Q[\rho]=\tilde Q[\rho+\delta\rho]-\tilde Q[\rho]$.
Here we use $\tilde Q[\rho]$ to distinguish from the discrete case.
\\

\subsection{The case  $0<q<1$}

The following theorem proves that, for $0<q<1$, instability only  can happen for singular distributions $\rho$.
In the theorem $||\obs||_\infty=\sup\{|\obs(x)|: x\in[0,1]\}$ denotes the so called supremum or uniform norm, which is just the smallest upper bound of $|\obs|$.
\begin{theorem}
Let $0<q<1$. Let $0<\rho$ be a non-singular probability distribution on
$I=[0,1]$. Let $G=\int_I dx\,\rho(x)^q$ and
let $0<\tilde\delta^q=\mu G/4$, for $0<\mu<1$, and $\delta\rho$ be a
variation of the distribution such that,
$\int_I dx |\delta\rho|=\delta\leq\tilde\delta$, and $0<\rho+\delta\rho$
is positive on $I$.
Furthermore, let $0<\obs$ be a strictly positive bounded observable on $I$,
then there exists a constant
$0<c<\infty$, such that
\begin{equation}
|\tilde Q[\rho]-\tilde Q[\rho+\delta\rho]|<c\delta^q\quad.
\end{equation}
Moreover, $c=
4G^{-2}||\obs||_{\infty}(1+||\obs||_\infty||\obs^{-1}||_\infty)/(1-\mu)$.
\label{theorem1}
\end{theorem}

The theorem states that, for positive bounded observables, $q$-expectation 
values are robust whenever
the distribution $\rho$ is non-singular.\footnote{When all considered $\rho$ are bounded by the same bound $0< \rho <B$, then the constant $c$ does 
not depend on the choice of $\rho$ and $\tilde Q$ is absolutely continuous on this domain.}  
The class of singular distributions is therefore the only class of distributions that contain all
possible violations to stability for  $0<q<1$, as long as the observable $\obs$ is bounded on
domain $I$.
The corresponding example in \cite{abe}  explicitly converges toward a singular distribution 
in the $W\to\infty$ continuum limit and thus violates  stability.

\subsection{The case $q>1$}

In contrast to the $0<q<1$ case,  instability for $q>1$ is not primarily 
due to singular distributions $\rho$, 
but due to the variation $\delta\rho$ having singular parts, i.e., due to an unbounded $\delta\rho$.
Note, that for bounded $\delta \rho$ to exist, $\rho$ also has to be non-singular. To keep 
$\int dx[ \rho(x)]^q$ and $\int dx [\rho(x)]^q \obs$
finite, we further restrict $\rho$ to be bounded.
\begin{theorem}
Let $q>1$ and let $m>0$ be an arbitrary but fixed constant.
Let $0<\rho$ be a probability distribution on $I=[0,1]$.
Let $\delta \rho$ be variations of $\rho$, i.e. $\rho+\delta\rho>0$.
Let $B>0$ be an arbitrary but fixed constant.
Let the variations $\delta\rho$ be uniformly bounded in the $m$-norm, i.e.
$||\delta\rho||_m<B$,
by this constant $B$. Further, let $||\delta\rho||_1=\delta$ and
let $0<\obs$ be a strictly positive bounded observable on $I$.
Let $\tilde\delta$ be an upper bound for the size of the variations
$\delta$ such that,
$(2^{1/q}-1)^{q/\gamma}(B^{q-\gamma}||\obs||_\infty||\obs^{-1}||_\infty)^{-1/\gamma} > \tilde\delta > 0$,
where $\gamma=(m-q)/(m-1)$,
then there exists a constant $0<R<\infty$, such that
\begin{equation}
|\tilde Q[\rho]-\tilde Q[\rho+\delta\rho]|<R\delta^{\gamma/q}\quad,
\end{equation}
and $R$ does not depend on the choice of $\rho$.
\label{theorem2}
\end{theorem}
Theorem 2 states that, for positive bounded observables, $q$-expectation 
values are robust whenever the distributions $\rho$ are uniformly  
bounded\footnote{When all considered distributions are bounded 
by the same upper bound  ($m\to\infty$ and $\gamma \to 1$) 
then $R$ can be chosen independent of $\rho$ and $\tilde Q$ becomes uniformly continuous}. 
Excluding unbounded variations from consideration therefore is sufficient to guarantee 
stability for the $q>1$ case in a general setting.
In the corresponding example in \cite{abe}, when formulated in the continuum limit, 
stability is violated by using unbounded variations.\footnote{The proof was carried out 
on the unit interval $I \in [0,1]$. This does not 
present a loss of generality, since the proofs can be extended to any bounded interval. 
For unbounded intervals, the proof gets more involved and requires 
to fix conditions that relate boundedness conditions of the observable  and the decay properties 
of $\rho$. }

Both theorems in \cite{hanel08}
are analytical statements about the continuity properties
of the $q$-expectation value as a non-linear functional without any
reference to thermodynamics.
They provide a useful and flexible mathematical tool to analyze
continuity properties for discrete sets as well as for the continuous case on $[0,\infty]$.

\section{Stability criteria for the $q$-expectation value for discrete finite probability functions}

We can now use the above theorems to prove the robustness of $q$-expectations
for bounded observables on finite discrete sets $i\in I_W\equiv\{1,2,\dots,W\}$. For this we map 
the discrete probability distribution onto the continuous interval $x\in[0,1]$ by identifying
probabilities $\{p_i\}_{i=1}^W$ with probability densities $\rho(x)=W p_i$ for $x\in[(i-1)/W,i/W]$, i.e.
step-functions on $[0,1]$. The observable $\{\obs_i\}_{i=1}^W$ gets identified with the step function
$f(x)=\obs_i$ for the associated interval $x\in[(i-1)/W,i/W]$.
Clearly, $0\leq\rho\leq W$ for all possible distributions of this kind and so are all possible variations
since $|\rho(x)-\rho'(x)|<W$. Since the observable $f(x)$ is bounded, 
all conditions needed for theorems 1 and 2 are met. 
Let $\gamma=q$ for $0<q<1$ and
$\gamma=1/q$ for $q>1$. For some given constant $0<\tilde \delta$
there exists a constant $C$, such that for all $|\rho-\rho'|<\delta<\tilde \delta$ it follows that 
$C\delta^\gamma>|\tilde Q[\rho]-\tilde Q[\rho']|$ where $\tilde Q[\rho]=\int_0^1 dx\,[\rho(x)]^q f(x)/\int_0^1 dx\, [\rho(x)]^q$
is the $q$-expectation for the continuous case.    
Now 
$\tilde Q[\rho]=\sum_{i=1}^W W^{-1} [W p_i]^q \obs_i/\sum_{i=1}^W W^{-1} [W p_i]^q = 
\sum_{i=1}^W p_i \obs_i/\sum_{i=1}^W p_i^q = Q[p]$, where $Q[p]$ is the $q$-expectation on $I_W$.
Moreover, $||\rho||_1=\int_0^1 dx\,|\rho(x)|=\sum_{i=1}^W |p_i|=||p||_1$ and
consequently one can pull back the result to the discrete case, which completes the proof.

\section{Comments on discrete infinite probability functions}

For discrete infinite distribution functions it was shown \cite{abe} that 
the $q$-expectation value is unstable. 
However, it is possible to state conditions under which robustness can be ensured.
This can be done, for instance, in the following way. 

Discrete finite sets  are closely related to continuous compact sets in the sense 
that  discrete sequences can be mapped into the compact interval with 
step functions, as discussed above.
In the same spirit,  discrete  infinite sets are intimately related to the
continuous unbounded set $[0,\infty]$, since again one can map the discrete 
infinite sequence into the continuous case in terms of step functions.
If one can find conditions which define classes of distribution
functions on $[0,\infty]$ that guarantee
continuity or absolute continuity of the functional, i.e., the
$q$-expectation, the same conditions are sufficient
for probabilities on infinite discrete sets.
Such classes can simply be derived  by using suitable differentiable
monotonous functions,
$g: [0,\infty]\to [0,1]$.  Let $g'$ denote the derivative of $g$ and
$g^{-1}$ the inverse function of $g$.
This maps the distribution function $\rho$, defined on $[0,\infty]$, to a
distribution function
$\tilde \rho(y)=\rho(g^{-1}(y))g'(g^{-1}(y))^{-1}$ on $[0,1]$
and also the
observable $\obs$ on $[0,\infty]$ gets mapped
to $\tilde\obs(y)=\obs(g^{-1}(y))g'(g^{-1}(y))^{q-1}$. Now one can apply
the conditions
used for the theorems 1 and 2 on $[0,1]$ and pull them back to
$[0,\infty]$.
It should be noted that different maps $g$ lead to different, yet
consistent boundedness conditions and decay properties
for observables $\obs$ and distributions $\rho$ on $[0,1]$.

To give an explicit example, let us consider the following problem. 
Suppose  we consider $\bar q$-exponential distributions of the form 
$\rho(x) \propto e_{\bar q }(-\beta x)\equiv [1-(1-\bar q)\beta x]^{\frac{1}{1-\bar q } }$
for $\bar q \ge 1$ and some $\beta>\beta_0$, and we want  its first $N$ moments 
 under the 
$q$-expectation,
\begin{equation}
\langle x^m \rangle_q\equiv \frac{\int dx [\rho (x)]^q x^m}{\int dx [\rho (x)]^q }
\label{ho}
\end{equation}
to be robust with respect to $g$, (i.e. $m\leq N$). For a more detailed discussion on 
moments under $q$-expectations, see \cite{tsallis08}.  
Assuming $q>1$, let us take $g(x)=1-1/(1+x)^\gamma$; then $g'(x)=\gamma (1+x)^{-\gamma-1}$. 
The   boundedness condition for the observables immediately  requires 
$\gamma>N/(q-1)-1$ and the decay property for the distributions implies 
$\bar q<1+1/(\gamma+1)$.
For $\bar q=1$ and  $m=1$ the example is the $q$-expectation value of the energy of the 
quantum harmonic oscillator,  
\begin{equation} 
 \langle E \rangle_q = \frac{\sum_{n=1}^{\infty }  n [e_1^{-\beta n }]^q   }{\sum_{n=1}^{\infty } [e_1^{-\beta n }]^q } 
         = \frac14 \sinh ^{-2} \left(  \frac{\beta q }{2} \right) \quad .
\end{equation} 
which shows continuity  in $\beta$ and robustness of the $q$-expectation 
under variations of the exponential distribution.\footnote{
In what concerns the use of continuous  distributions for calculating entropies and similar 
quantities, such as Eq. (\ref{ho}),   the reader must be aware of a relatively well known difficulty. 
If we make a change of variables $y=f(x)$, say in Eq. (\ref{ho}), we immediately see that the result is 
not invariant. In the spirit of Kullback and Leibler entropy \cite{kull} the problem is easily resolved 
in terms of a {\em reference distribution} $r(x)$, which, except  at infinity, nowhere vanishes. 
For example, for $q>0$,  the quantity in  Eq. (\ref{ho}) would be replaced by 
$\frac{\int dx  r(x)  [\rho (x)/r(x)]^{q} x^m}{\int dx  r(x) [\rho (x)/r(x)]^{q} }  $.
}

\section{Robustness of R\'enyi entropy  for continuous and discrete finite 
distribution functions}

Here it suffices to discuss the case $0<q<1$ since for $q>1$ R\'enyi entropy 
\begin{equation} 
 S_q^{\rm R} =\frac{\ln  \sum_{i=1}^W p_i^q }{1-q}
\end{equation}
strictly speaking ceases to be a proper entropy because it is not concave.
Substituting the probabilities $p_i$ by step-functions 
$\rho^{(W)}(x)=W p_i\leq W$ for $x\in[(i-1)/W, i/W]$
which represent the discrete probability $p_i$ as a probability density on
$[0,1]$, we get 
\begin{eqnarray} 
 S_q^{\rm R}
              &=& \ln  W  + \frac{ \ln \int_0^1 dx\, [\rho^{(W)}(x)]^q }{1-q} \quad.
\end{eqnarray}

Boundedness of the distributions allows  to use  propositions
(4), (6) and (8) in \cite{hanel08} which have been used to prove theorem (1).
Note that, due to the upper bound $W$, it follows that $\int dx \rho(x)\geq
W^{q-1}$.
Let $\tilde\delta=\mu W^{q-1}/4$ and $||\delta\rho||_1=\delta<\tilde\delta$, as  in \cite{hanel08}.
Now we get
\begin{equation}
 |S_q^{\rm R}[\rho]-S_q^{\rm R}[\rho+\delta \rho]|= \left| \ln \frac{ \int_0^1 dx\, [\rho(x)]^q }{  \int_0^1 dx\, [\rho(x)+\delta\rho(x)]^q } \right|
\quad .
\end{equation}
Using propositions (4) and (6) from \cite{hanel08} one finds 
$1-a\delta^q  < |\int_0^1 dx\, \rho(x)^q/\int_0^1 dx\,
(\rho(x)+\delta\rho(x))^q|<1+a\delta^q$
and $a=4W^{1-q}/(1-\mu)$. Since $|\ln (1+x)|<2|x|$, for $|x|\ll1$  it
follows that for sufficiently small $\delta$,  
\begin{equation} 
 |S_q^{\rm R}[\rho]-S_q^{\rm R}[\rho+\delta \rho]|< 2a \delta^q \quad .
\end{equation}
This shows both the uniform continuity of the continuous R\'enyi entropy
for the class of uniformly bounded probability distributions in $L_1([0,1])$, 
and the absolute continuity of R\'enyi entropy
for probabilities on finite sets. Similar arguments show robustness also for $q>1$, 
despite lack of concavity of $S_q^{\rm R}$. 

\section{Robustness of the entropy $S_q$ for continuous and discrete finite 
distribution functions}

One can prove robustness of the nonadditive entropy 
\begin{equation}
 S_q = \frac{1-\sum_{i=1}^{W}  p_i^q}{q-1} 
\end{equation}
 for discrete finite sets by mapping
the probabilities $\{p_i\}_{i=1}^W$ onto a distribution on $\rho$ on
$[0,1]$ by step functions, as above. Since all step functions $\rho$ that
represent some $\{p_i\}_{i=1}^W$ are bounded by $W$, it is sufficient
to prove robustness for the continuous case. 
$\rho$ is from a uniformly bounded class of distribution
functions on $[0,1]$, i.e., there is a constant $W>0$ such that all $\rho$  are bounded by $W$. 
Consider variations $\delta\rho=\rho'-\rho$ 
such that both $\rho$ and $\rho'$ are bounded by $W$ and $||\delta\rho||_1=\delta$, where $\delta$ is 
sufficiently small. 
Now, $|S_q[\rho]-S_q[\rho+\delta\rho]|=|q-1|^{-1}|\int_0^1 dx\, \rho^q - \int_0^1 dx\, (\rho+\delta\rho)^q|$.
It therefore is an immediate consequence of propositions (6) and (4)
in \cite{hanel08},  that for $0<q<1$, 
$|S_q[\rho]-S_q[\rho+\delta\rho]|<4\delta^q$. For $q>1$ we use
propositions (11) and (13) to find $|S_q[\rho]-S_q[\rho+\delta\rho]|<R\delta^{\frac1q}$. 

In what concerns discrete  infinite distribution functions, $S_q$ has been shown to be Lesche-stable  
\cite{abe00}.

\begin{table*}
\begin{center}
\begin{tabular}{|c|c|c|c|}
\hline
  & $S_q$,  for $q>0$
  & $S_q^{\rm R}$, for $q>0$ 
  &  $q$-expectation value, for $q>0$ \\
\hline
 \begin{tabular}{c} bounded continuous \\ (physically admissible)  \end{tabular} 
 & robust 
 & robust 
 & robust \cite{hanel08} \\
\hline
 \begin{tabular}{c} discrete finite \\ ($W$ finite) \end{tabular} 
 & robust
 & robust
 & robust  \\ 
\hline
 \begin{tabular}{c} discrete infinite \\ ($\lim W \to \infty $) \end{tabular} 
 & Lesche-stable \cite{abe00}
 & \begin{tabular}{c} Lesche-unstable \cite{lesche}; robust \\   for typical physical cases \end{tabular}
 & \begin{tabular}{c} unstable \cite{abe}; robust for  \\   typical physical cases \end{tabular} \\
\hline
\end{tabular}
\caption{Table of stability/Lesche-stability and robustness for the functionals $S_q$, $S_q^{\rm R} $  and the 
$q$-expectation value for the underlying nature of the distribution function, i.e.  continuous, 
discrete finite or discrete infinite.  In the continuum, admissible means that a
non-negative entropy   is required. 
Boundedness automatically guarantees robustness \cite{hanel08}. 
The term {\em Lesche-stable} is used for infinite distribution functions 
when Eq. (\ref{lesche}) holds, {\em stable} is used when there is no division by a maximum taken in  Eq. (\ref{lesche}). 
{\em Robustness} is used for bounded distributions in the continuum and for discrete finite probability functions.
For the  $q$-expectations for the discrete infinite case, robustness is understood under the decay properties of the distributions, 
as specified in the text.
Lesche-stability and stability are sufficient but not necessary for robustness. 
$S_{\rm BG }$ and standard expectation values  ($q=1$) are Lesche-stable and stable, respectively. 
 }
\end{center}
\end{table*}

\section{Conclusion}\label{discussion}

To summarize, we discussed  the concept of physical robustness in contrast to the more restrictive 
mathematical stability of thermostatistical functionals under variations of their 
underlying distribution functions. We argue that while several important functionals, 
such as the R\'enyi entropy or the $q$-expectation value are unstable in the strict sense, 
restriction to physically relevant distribution functions, ensures robustness of these functionals. 
For a distribution to be physically relevant we require that its associated entropy be non-negative. We further restrict to distributions which are bounded in the continuum. This excludes, for example, distributions involving 
Dirac deltas. 
Our results are summarized in Table 1, where we indicate the type of stability, Lesche-stability or 
robustness for the functionals $S_q$, R\'enyi entropy $S_q^{\rm R}$ and the $q$-expectation value, 
for the paradigmatic types of distribution functions -- continuum, discrete finite, and discrete infinite 
-- that we have focused on here. 
The term Lesche-stable is  used when Eq. (\ref{lesche}) holds,  stable refers to the situation where 
no division by a maximum is taken in   Eq. (\ref{lesche}), and  robustness -- in the above-defined sense -- 
is found for admissible (non-negative entropy) and bounded distributions in the continuum 
and for discrete finite probability functions.
In the case for discrete infinite distribution functions which are known to cause instabilities 
for some functionals \cite{lesche,abe},
one can show that there exist  paradigmatic robust examples once decay properties of the distributions are specified. 
In this context one can show that systems such as the harmonic oscillator 
are robust under variations of the (inverse) temperature.  
One can of course  think of physical distribution functions which have positive BG entropy 
but are unbounded, such as e.g. power-law divergences. These cases remain to be discussed
but exceed the present scope.

We conclude by  stating that the concept of stability might be overly strict for 
physical applications. 
This is in accordance to conclusions drawn in \cite{jizbaarimitsu04}.
When limited to the class of physically admissible and bounded distribution 
functions, it is conceivable 
that physical robustness of virtually all thermodynamic functionals will be guaranteed.
 
We acknowledge  fruitful remarks by E.M.F. Curado and C. Beck.  
R.H. and S.T. acknowledge  great hospitality at CBPF. 
This work was   partially supported by Faperj and CNPq (Brazilian agencies),
and Austrian Science Fund FWF Project P19132.

\end{document}